# Evaluation Process for an Introductory Programming Course Using Blended Learning

Francisco de Assis Zampirolli, Federal University of ABC, Santo André, Brazil

Denise Hideko Goya, Federal University of ABC, Santo André, Brazil

Edson Pinheiro Pimentel, Federal University of ABC, Santo André, Brazil

Guiou Kobayashi, Federal University of ABC, Santo André, Brazil

[Paper in translation process for English and publication]

## ABSTRACT

A oferta de um mesmo curso simultaneamente para milhares de alunos, seja no ensino presencial ou no ensino a distância, requer que sejam adotadas uniformizações a fim de permitir a comparação de desempenho entre turmas e avaliar o sucesso ou não dos processos de ensino-aprendizagem. Este trabalho tem por objetivo apresentar um Evaluation Process (EP) desenhado para um curso de Introdução à Programação (IP) para alunos do Bacharelado em Ciência e Tecnologia da Universidade Federal do ABC. Inicialmente apresenta-se o cenário do curso na modalidade presencial (IP-Class), ofertado anualmente para cerca de 2000 alunos, apontando os problemas identificados e, em seguida, descreve-se o EP desenhado e aplicado em turmas Blended Learning (IP-BL), incluindo o uso de ferramentas de software desenvolvidas para esse fim. Os dois cenários (IP-Class e IP-BL) são comparados com o uso de dados de todas as turmas dos últimos 9 anos de aplicação do curso.

## 1. INTRODUCTION

Pessoas com conhecimentos em computação estão mais preparadas para acompanhar os avanços tecnológicos que estamos imersos nos dias atuais. Além disso, a computação é uma área ampla e se conecta com várias outras áreas de forma interdisciplinar (ACM-IEEE, 2013).

Nesse sentido, o projeto pedagógico da Universidade Federal do ABC (UFABC) considera que todos os alunos que ingressam na universidade, por meio do Bacharelado em Ciência e Tecnologia, devem cursar uma disciplina introdutória de programação de computadores.

A UFABC é uma universidade nova, tendo iniciado a primeira turma no segundo semestre de 2006 e possui um projeto pedagógico inovador, inspirado no Processo de Bolonha (Keeling, 2006), no qual o aluno ingressa necessariamente em um bacharelado interdisciplinar (BI), que tem a duração de 3 anos. A UFABC tem atualmente dois BIs: Bacharelado em Ciência e Tecnologia (BCT) ou Bacharelado em Ciências e Humanidades (BCH). Em 2017 foram 1560 alunos ingressantes no BCT e 400 alunos ingressantes no BCH, com um ingresso por ano. Enquanto cursa o BI, o aluno pode escolher cursar disciplinas específicas de outros cursos a sua escolha. Atualmente, a UFABC oferece 8 engenharias, 13 bacharelados e 5 licenciaturas. Assim, em 4 anos, o aluno pode concluir um BI e um outro bacharelado como o Ciência da Computação, ou em 5 anos, pode concluir um BI e uma engenharia.

O ano letivo da UFABC possui 3 quadrimestres, e cada disciplina tem a duração de 12 semanas. O curso de Introdução à Programação (IP) da UFABC é a disciplina Information Processing, oferecida a todos os alunos do BCT. O plano de ensino de IP é dividida em temas teóricos e atividades em laboratório, e contém: *Introdução a algoritmos; Variáveis e tipos de dados; Operadores aritméticos, lógicos e precedência; Métodos/Funções e parâmetros; Estruturas de seleção; Estruturas de repetição; Vetores; Matrizes; Entrada e saída de dados; Depuração; e Melhores práticas de programação.*

Desde o projeto pedagógico do BCT de 2009, o curso de IP é ofertado no terceiro quadrimestre, final do primeiro ano de ingresso do aluno, ou seja, num momento em que o aluno ainda não escolheu no que pretende se graduar, além do BCT. Pode ser que ele esteja, por exemplo, dirigindo sua formação para o Bacharelado em Biologia ou para o Bacharelado em Ciência da Computação. Mesmo assim, ele deverá cursar IP por ser uma disciplina obrigatória do BCT. Assim, a variedade de perfis e a grande quantidade de alunos exigem um grande empenho dos professores do curso para manter um padrão de qualidade no ensino e na avaliação.

Na modalidade presencial, IP é executada em 12 semanas consecutivas com 5 horas aulas semanais, sendo 3 horas em salas de teoria e 2 horas em laboratórios computacionais. As turmas presenciais (IP-Class) são organizadas da seguinte forma: turmas de teoria com cerca de 90 alunos (um professor) por turma, e estes mesmos 90 alunos são distribuídos em 3 turmas de laboratório com 30 alunos cada (três professores distintos).

Considerando-se que 1560 alunos devem cursar IP em cada ano (sem contar as turmas extras para atender as reprovações), normalmente são preparadas 18 turmas de teoria (90 alunos cada) e 52 turmas de prática (30 alunos) podendo envolver até 70 professores no total. Apesar de haver uma coordenação de disciplina (atribuída a um dos professores) a cada período em que o curso é ofertado, não tem sido adotado um padrão de avaliação obrigatório, como por exemplo a aplicação de provas unificadas, o que resulta em discrepâncias e variações nos resultados das avaliações. Naturalmente, apenas a aplicação de provas unificadas não seria suficiente para eliminar as discrepâncias, mas poderia contribuir para a identificação de problemas no processo de ensino-aprendizagem, onde dados consolidados de diversas turmas poderiam apresentar desempenhos muito diferentes.

A partir de 2013, a UFABC passou a oferecer também turmas de IP na modalidade Blended Learning (IP-BL), cujas aulas e as atividades são desenvolvidas a distância via Internet, sendo as avaliações, como provas e projetos, realizadas presencialmente. O objetivo inicial da modalidade Blended Learning (BL) foi diminuir a grande demanda de professores e de infraestrutura (laboratórios computacionais) exigidas para as turmas de IP-Class (Braga et al., 2014) e ampliar a oferta de vagas. Esta modalidade tem sido ofertada de maneira consistente desde então, com uma média de 180 vagas por quadrimestre. As aulas são realizadas com o suporte de um Ambiente Virtual de Aprendizagem (VLE) em que todos os alunos de IP-BL de um quadrimestre são cadastrados em uma única turma, bem como todos os professores (cerca de 5) e tutores (cerca de 7). Essa concentração dos alunos, professores e tutores num único espaço virtual (sala de aula) propicia um padrão em relação a conteúdo, material instrucional, comunicação, feedback, etc. Num comparativo com o sistema presencial é como se todas as turmas e professores estivessem numa única sala. O cenário de IP-BL (sala virtual única) criou portanto a possibilidade de se desenhar um Evaluation Process (EP).

Este trabalho tem por objetivo apresentar um EP desenhado para a disciplina Information Processing (IP) para alunos do Bacharelado em Ciência e Tecnologia da Universidade Federal do ABC. O EP foi utilizado no contexto de turmas na modalidade Blended Learning (IP-BL).

Este artigo está organizado como segue: a Seção 2 apresenta uma breve revisão de literatura e sobre trabalhos que fazem uso das tecnologias de informação e comunicação no apoio ao ensino de programação; a Seção 3 apresenta o cenário do curso na modalidade presencial (IP-Class) apontando os problemas identificados; a Seção 4 descreve-se o EP desenhado e sua aplicação em turmas Blended Learning (IP-BL) tentando cercar os problemas identificados, e também descreve algumas ferramentas desenvolvidas para apoiar a aplicação do EP no contexto de IP-BL; a Seção 5 discute os resultados e, por fim, a Seção 6 apresenta as considerações finais e trabalhos futuros.

## 2. REVISÃO DE LITERATURA

Esta seção tem por objetivo apresentar uma breve revisão de literatura sobre trabalhos relacionados a ensino de programação na modalidade a distância ou blended learning, especialmente os que fazem uso de tecnologias da informação e comunicação para auxiliar o EP.

### 2.1. Cursos presenciais versus cursos online

Em Castillo-Merino & Serradell-López (2014), foi constatado que não existe uma diferença significativa no desempenho de alunos que cursam presencialmente ou online. Porém a motivação do aluno é o fator que mais influencia positivamente no aprendizado online. Nesse trabalho também foi constatado que alunos mais velhos e com experiências em *Information and Communication Technologies* (ICTs) apresentam desempenhos melhores em cursos online. O trabalho faz um levantamento bibliográfico extenso e destaca que ainda existe resistência dos professores tradicionais a adotarem o modelo de ensino a distância.

García et al. (2014) apresenta uma correlação existente entre os desempenhos das atividades (com peso de 40% do resultado final) desenvolvidas pelos alunos durante um curso online e o resultado do exame final presencial (60%). O curso é de Economia Estatística para o Business Administration and Management degree, com partes teóricas e práticas usando planilhas eletrônicas. García et al. (2014) sugere aumentar o número de atividades (nos testes empíricos, foram 5 atividades), incentivando o uso das ICTs. Ainda, foi constatado que quando os alunos obtêm resultados ruins nas primeiras atividades, eles tendem a deixar o curso.

Brito & De Sá-Soares, (2014) realizaram um experimento no curso de primeiro ano da universidade em fundamentos da programação de computadores, e indicaram que é necessário aumentar a frequência de avaliações. Além disso recomendaram dar retornos regulares dos desempenhos dos alunos, pois quando estes entendem que não conseguem resolver um problema de programação, geralmente é tarde demais para recuperar uma deficiência ocasionando uma desistência do curso.

Djenic et al. (2011) descreveu um curso blended learning, utilizando o Learning Management System (LMS) Moodle (moodle.com) para disponibilizar conteúdos eletrônicos: livro eletrônico, animações em Flash, ambiente de programação (Integrated Development Environment – IDE) e quiz (testes de múltipla-escolha). A implementação do curso foi com 50% de aulas presenciais e 50% a distância, em dois cursos de programação nos dois primeiros anos da graduação. O aluno podia escolher entre o modelo tradicional de ensino e o blended learning e a aceitação deste segundo foi ascendente nos anos de 2007 e 2008: 18%, 29%, 40% e 53%, respectivamente. Nos experimentos, 147 alunos acompanharam o modelo blended learning, dos 420 que participaram de ambos os tipos de cursos. Nestes experimentos 58% foram aprovados no modelo blended learning, contra 55% no modelo tradicional.

## 2.2. Ferramentas de apoio ao ensino de programação

Gusev et al. (2016) apresentou o uso de *Information and Communication Technology* (ICT) para avaliar alunos utilizando o e-testing, uma ferramenta de testes eletrônicos de múltipla escolha tradicional e também uma versão com imagens interativas, onde o aluno clica sobre uma imagem para escolher a alternativa desejada. O banco de questões é agrupado por conteúdos. Se três questões sobre um mesmo conteúdo forem respondidas corretamente, segue para o próximo conteúdo, caso contrário, outras questões do mesmo conteúdo são fornecidas ao aluno para reforçar o aprendizado. São fornecidas no banco de 6 a 10 questões envolvendo um mesmo conteúdo, com o mesmo nível de dificuldade. O foco do artigo está no banco de questões e pode ser construído usando arquivos XML com muitos tags para descrever as várias variações de questões de múltipla-escolha, como: pesos, número de respostas corretas, e quanto penalizar se a resposta for errada. Como geralmente somente programadores conseguem escrever as questões usando o formato XML, foi desenvolvida também uma macro simplificada no Word para converter um documento Word para o formato XML, a ser lido pela ICT. As questões com imagens interativas são fornecidas pela estrutura XML com tags específicas: *image_file*, *image_resolutions*, *XYpar_correct*, etc. Este trabalho é semelhante ao apresentado em Zampirolli et al. (2016) e adaptado neste artigo, porém, com a diferença de elaborar questões em papel e utilizar a sintaxe do LaTeX (miktex.org), no lugar do XML.

Como propostas de novas ICTs, em Del Fatto et al. (2016) foi apresentada a ferramenta chamada eXtreme Apprenticeship (XA) aplicada na disciplina de Sistemas Operacionais em um curso de mestrado utilizando programação em Bash (comandos do Shell de um Sistema Operacional), desenvolvida no terminal do Unix. Foram ao todo 45 exercícios Bash enviados através do LMS Moodle durante 6 semanas e corrigidos de forma automática. Estes exercícios contabilizaram 25% da nota final e o aluno tinha que entregar pelo menos 70% destas atividades semanais, podendo submeter várias vezes o mesmo exercício até obter sucesso (acerto) ou atingir o limite de submissões (8 tentativas). Um projeto presencial e individual também foi aplicado em laboratório, sendo supervisionado por um assistente, valendo 25% na nota final. Os outros 50% da nota final foram obtidos através de uma prova conceitual escrita. O experimento foi realizado em uma turma de 28 alunos, com 5 desistentes. Os 23 alunos restantes fizeram todas as atividades, com 4 reprovações. O problema destes exercícios Bash, e outros usando juízes online, como BOCA (goo.gl/kHDVfy), URI (urionlinejudge.com.br), repl.it, entre outros (Toledo & Moron, 2014), é que a pontuação ocorre somente quando o exercício está correto, se tornando rigoroso e desestimulante para o aluno em um curso introdutório de programação.

Em Kose & Deperlioglu (2012) foi apresentado uma ICT utilizada na Universidade Afyon Kocatepe na Turquia que inclui ferramentas padrão de e-learning e dois subsistemas, CTutor e ITest. O primeiro é destinado a resolução de problemas de programação na linguagem C, que diagnostica o nível de conhecimento do aluno e elabora questões específicas, fornecendo feedbacks e dicas para ajudar a resolver o problema. No CTutor foi desenvolvida uma interface que permite ao aluno usar mecanismo de clicar e soltar para compor o código, porém incorporado a um sistema de e-learning. As mensagens de erro usam a modelagem baseada em restrições, ou seja, o aluno aprende com os erros. Existem dois tipos de problemas a serem resolvidos pelos alunos: (a) usando apenas instruções sequenciais e (b) usando blocos de instruções, como estruturas condicionais e laços. As escolhas de novos problemas para os alunos são realizadas conforme o tempo de resolução e os problemas já resolvidos pelo aluno. Já o ITest fornece um ambiente de avaliação no qual os alunos podem fazer testes de múltipla escolha preparados de acordo com o seu nível de aprendizagem. O curso de programação C foi dividido em dois semestres, com 14 semanas cada e 60 alunos por turma. No início de cada semestre o professor apresenta de forma presencial o conteúdo necessário para o aluno poder seguir o aprendizado no LMS (neste trabalho não foi especificado o número de horas). Depois, o aluno continua o curso pelo ambiente virtual, e na 7ª semana o aluno faz um exame não presencial, valendo 20% na nota final. Na 14ª semana o aluno faz um exame presencial valendo 80% na nota final. Estes exames são realizados no próprio sistema LMS utilizando os bancos de questões disponíveis. Este sistema não suporta ainda ferramentas de Inteligência Artificial para as escolhas de exercícios de cada aluno.

## 3. CONTEXTO DO CURSO PRESENCIAL: IDENTIFICANDO PROBLEMAS

Esta seção tem por objetivo apresentar o cenário do curso de IP-Class identificando os principais problemas relacionados ao grande número de turmas e de professores, além dos problemas decorrentes da ausência de avaliações unificados.

### 3.1. Números de IP-Class

Anualmente, ingressam 1560 alunos no BCT. Como o projeto pedagógico da UFABC permite que o aluno construa sua trajetória, nem todos esses alunos se matriculam simultaneamente nas mesmas disciplinas.

Em 2017.1, 1203 alunos se matricularam em IP-Class, distribuídos em 40 turmas práticas e em 13 turmas teóricas. Essa grande demanda de alunos se torna um problema na gestão de ensino-aprendizagem, agravados também pelos diversos perfis de alunos e professores. A avaliação desses alunos de forma justa também é um grande desafio para os professores.

### 3.2. Problemas identificados no IP-Class

Algumas dificuldades encontradas em IP-Class são descritas a seguir:

**Conteúdo:** Uma dificuldade em IP-Class é gerenciar todos os professores para unificar o conteúdo apresentado ao aluno. Apesar de haver uma recomendação de linguagem de programação a ser adotada por todos, infelizmente os professores acabam utilizando linguagens e processos de ensino-aprendizagem distintos, conforme a sua afinidade para abordar o plano de ensino do curso IP.

**Critério de avaliação:** Uma vez que cada professor pode usar critérios distintos de avaliação, os dados mostram resultados discrepantes entre as turmas. Por exemplo, das 391 turmas de IP-Class ofertadas entre 2009.3 e 2017.2, 16 turmas (ou 4,09%) tiveram 100% de aprovação dos alunos. Num outro extremo, houve duas turmas, uma de 12 alunos e outra de 13 alunos, em que apenas um único aluno foi aprovado em cada turma (menos de 8,33% de aprovação). Essas turmas deveriam ter 30 alunos em média no início do quadrimestre e possivelmente houve cancelamentos de matrículas.

**Escolha de turma e cancelamento:** Até 2015, o aluno podia cancelar várias disciplinas no primeiro mês do quadrimestre e isso gerava excesso de cancelamentos. Desde 2015, o aluno tem direito de cancelar somente uma disciplina após cursar duas semanas de aula, não contabilizando esta desistência no sistema acadêmico. No atual sistema de matrícula, normalmente disponibiliza-se o nome do professor que ministrará a disciplina, podendo o aluno optar por não se matricular com um determinado professor. Associada à avaliação despadronizada e não unificada, o aluno acaba muitas vezes desistindo de cursar uma disciplina com um determinado professor. Destaca-se ainda que os alunos têm acesso a um aplicativo próprio, chamado UFABCHelp (ufabchelp.me), no qual avaliam livremente os professores: alguns professores são classificado como mais "fáceis" e outros como mais "difíceis". Essas avaliações muitas vezes usadas pelos alunos para escolher ou não a turma de um de determinado professor e em tese, minimizam os cancelamentos. No entanto, supondo, de forma hipotética e extrema, que um aluno consiga sempre escolher os professores "fáceis"" ou menos exigentes, ele poderá chegar ao mercado de trabalho com uma formação deficiente, prejudicando inclusive o nome da instituição que o formou. Assim, a ausência de padrões de avaliação propiciam que numa mesma instituição haja profissionais formados, com níveis muito diferentes. Isso pode ser visto, por exemplo, dentre 391 turmas de IP-Class, 7 tinham em média 15 alunos, onde cerca de 50% dos alunos cancelaram possivelmente por conta de um professor mais rigoroso, e onerando a instituição. Nessas 7 turmas, o índice de reprovação foi de 68.26%, bem superior à média de 30.33% de todas as 391 turmas de IP-Class. Por outro lado, das 16 turmas com 100% de aprovação, a média de alunos por turma foi de 27 alunos, e a média de alunos com conceito máximo ("A") foi de 13 alunos por turma. Será que essas turmas com extremos de reprovações e aprovações tiveram os mesmos critérios de ensino e de avaliação?

**Conceito:** Os conceitos finais de cada aluno devem ser lançados no sistema acadêmico da UFABC como letras A, B, C e D (sendo A o melhor conceito), para os aprovados, "F" para os reprovados por baixo desempenho e "O" para os reprovados por falta (mais de 25% de ausências nas aulas). Quando o aluno obtém conceito "O", limita o número de reprovações em uma mesma disciplina e também a possibilidade de conseguir bolsas de auxílio moradia, monitoria e de iniciação científica.

Em IP-Class o professor de prática gera um conceito para o aluno (peso de 40%) e o professor de teoria gera um outro conceito (peso de 60%). Esses dois conceitos precisam ser unificados para gerar o conceito final do aluno na disciplina. O fato dessas avaliações, nas duas partes da disciplina, serem independentes (e sem uniformidade) é mais um elemento de dificuldade.

**Recuperação:** Por uma resolução interna da instituição (goo.gl/QRGaTd, em português), a partir de 2015, o aluno que ficar com conceito final D ou F tem o direito a um mecanismo de recuperação (REC). Como a resolução não prevê uma regra de cálculo do conceito final após a recuperação, alguns professores consideram o valor da prova de recuperação como o conceito final, outros consideram a maior nota entre a prova de recuperação e o conceito anterior, e outros ainda consideram a média das notas, entre outros critérios. Observem que cada um desses critérios adotados de forma individual por um professor pode melhorar ou não a média de reprovações e de conceitos finais de uma turma.

**Reprovação por ausência:** no ensino presencial (goo.gl/kjvkMo, em português), o aluno que não participar de pelo menos 75% das aulas deve ficar com conceito O, sem direito a REC. Em Blended Learning (goo.gl/sTrJBR, em português), se o aluno deixar de entregar pelo menos 75% das atividades, deve ficar com o conceito final F, mas não está claro se esse aluno tem o direito de fazer a REC (também por lei, cabe à IES definir critérios de compensação de ausências não justificadas). Essas diferenças exercem influência na avaliação final.

Alguns dos problemas e soluções relatados neste artigo dizem respeito ao que ocorre na UFABC. Porém, problemas semelhantes (como exemplo: muitos alunos em turmas diferentes, falta de interação entre os professores, ausência de padrões de avaliações, entre outros) podem ocorrer em outras instituições. Assim, é válido analisarmos esses problemas, descrever as soluções utilizadas, e cada instituição poderá adotar e adaptá-las caso avalie como interessantes.

A Tabela 1 apresenta um resumo dos problemas identificados em IP-Class e a última coluna apresenta as ações para contornar esses problemas a partir de um modelo Blended Learning (IP-BL), cujos detalhes serão apresentados e discutidos nas próximas seções.

*Table 1. Problemas identificados nas turmas de IP-Class e soluções utilizando IP-BL.*

| **Problema** | **IP-Class** | **IP-BL** |
|---|---|---|
| **Conteúdo** | Não unificado. Como verificar se todos os professores estão ensinando o mesmo conteúdo? | Utilizar o mesmo material instrucional (textos e videoaulas) disponibilizados no VLE . |
| **Critério de avaliação** | Não unificado na teoria e na prática. Cada professor usa critérios próprios de avaliação. | Definir um critério único (número de atividades, critérios de correção, pesos etc). |
| **Escolha de turma e cancelamento** | Alunos com melhores conceitos têm mais chance de escolher turmas com professores não rigorosos, gerando um ciclo vicioso. | Alunos se matriculam numa única turma e seu professor responsável é designado posteriormente. |
| **Conceito** | Diversidade na composição e divulgação dos conceitos de teoria e de prática. | Unificado: acompanhamento por meio de planilha e com uso de scripts desenvolvidos. |
| **Recuperação** | Quem e quando aplicar a prova de recuperação (teoria ou prática)? Como | Unificado. |
| | compor o conceito final, após REC? | |
| **Reprovação por ausência** | Não existe controle rigoroso de frequência do aluno e compensação de ausências não justificadas do aluno. | Controle de frequência definido pelas atividades entregues no VLE. Abono de ausências com a REC. |

Braga et al. (2014) apresentam o processo de ensino-aprendizagem do curso de IP-BL ofertado no primeiro experimento em 2013, como material didático utilizado, tutores, professores e a avaliação de satisfação dos alunos com relação a esta nova modalidade de IP-BL. Agora, neste artigo, vamos detalhar as melhorias realizadas nestes últimos 5 anos de ofertas de IP-BL, principalmente no que diz respeito ao Evaluation Process (EP).

## 4. PROPOSED EVALUATION PROCESS FOR IP-BL

Esta seção tem por objetivo descrever o Evaluation Process (EP) desenhado e sua aplicação nas três últimas turmas de Blended Learning (IP-BL) tentando cercar os problemas identificados na seção anterior, e também descreve algumas ferramentas desenvolvidas para apoiar a aplicação do EP no contexto de IP-BL.

### 4.1. Sala virtual unificada

Atualmente a UFABC utiliza um Ambiente Virtual de Aprendizagem (VLE) denominado Tidia-ae (tidia4.ufabc.edu.br) (Beder et al., 2007). O Tidia-ae é baseado no projeto Sakai (sakaiproject.org) e foi adaptado com financiamento da agência de fomento do estado de São Paulo - FAPESP e implantado em várias instituições de ensino do estado de São Paulo.

#### *4.1.1. Conteúdo e estratégias comuns*

O Tidia-ae é usado para distribuir o material didático, troca de mensagens entre alunos, professores e tutores e receber as atividades e as avaliações presenciais em laboratório dos alunos.

As atividades do curso de IP-BL são divididas em 12 semanas consecutivas. Cada uma das 8 primeiras semanas encerram-se com avaliações formativas (formative assessment). A cada semana, os alunos entregam dois ou três testes de retorno automático. Também, em cada semana os alunos entregam 1, 2 ou 3 listas de exercícios (contendo textos ou códigos de programas). Totalizando 17 testes e 19 listas. Estas listas de exercícios são corrigidas manualmente pelos tutores.

Infelizmente o Tidia-ae é pouco flexível para a realização de cálculos para a atribuição de conceitos finais de um aluno. Desta forma, se faz necessário usar uma planilha eletrônica compartilhada entre os professores no Google Drive (drive.google.com).

### *4.1.2. Número equilibrado de alunos e tarefas por professor e tutor*

No início de cada quadrimestre, são ofertadas duas turmas de IP-BL, uma no campus de Santo André (SA) e uma outra no campus de São Bernardo do Campo (SBC), na região metropolitana de São Paulo. Porém, em cada quadrimestre existe uma única turma (virtual) no VLE para todos os alunos de SA e SBC. No segundo quadrimestre de 2017 (2017.2) matricularam-se 107 alunos em SA e 54 alunos em SBC.

Cada tutor acompanha e corrige atividades de cerca de 25 alunos durante todo o curso. Em IP-BL, como se trata de uma turma unificada (virtual), é possível distribuir os alunos entre os professores de forma equânime, sem consumir mais recursos humanos. Assim, cada professor acompanha 10 alunos por crédito (em 2017.2 foram 30 a 40 alunos por professor em IP-BL). Diferentemente do modelo presencial, no qual o professor de teoria chega a acompanhar até 90 alunos (3 créditos) e cada professor de laboratório até 40 alunos (2 créditos).

## 4.2. Avaliação da Aprendizagem

Pelas leis brasileiras (goo.gl/kjvkMo, em português), a maior parte do conceito final de um aluno deve ser atribuído por avaliações presenciais e o aluno deve entregar pelo menos 75% das atividades em um curso a distância (goo.gl/sTrJBR, em português). O aluno tem apenas 4 encontros presenciais obrigatórios: um na primeira aula, na qual o professor apresenta o ambiente virtual, o plano de ensino e os critérios de avaliação; os outros três encontros obrigatórios são para as avaliações Exam1, Project e Exam2, descritas a seguir.

### *4.2.1. Critérios de Avaliação*

Nos dois primeiros quadrimestres de 2017, foi adotada a seguinte distribuição de pesos avaliativos para o curso de IP-BL:

**Activities:** 10% - atividades semanais, com os testes e listas que compõem o conjunto de avaliações formativas semanais. Levando em consideração que essas atividades semanais têm o propósito principal de avaliar a participação e o aprendizado de cada tópico do aluno no curso, não são atribuídos uma porcentagem alta do conceito final. De forma comparativa, não podemos aprovar os alunos que apenas se esforçaram indo nas aulas presenciais, mas não conseguiram ter conceitos suficientes nas provas;

**Exam1:** 35% - primeira avaliação presencial realizada após a 3ª semana de aulas. Essa é a primeira avaliação somativa do curso e abrange os tópicos iniciais da disciplina, que são estruturas sequenciais, estruturas de seleção e laços de repetição. O Exam1 inclui 3 questões dissertativas;

**Project:** 10% - atividade presencial em grupo de até 4 alunos, realizada na 7ª semana, em laboratório. Esse projeto, por ser uma atividade em grupo, também é atribuído um peso baixo para o cálculo do conceito final. Pois, não se sabe se todos os integrantes do grupo realmente conseguem reproduzir os experimentos (Brito & De Sá-Soares, 2014), servindo mais como uma preparação para o Exam2, em laboratório. Os códigos produzidos neste projeto devem ser anexados em atividades específicas no VLE;

**Exam2:** 45% - segunda avaliação somativa e presencial, realizada em laboratório, na 10ª semana, envolvendo todo o conteúdo do curso. Os códigos produzidos nesta prova devem ser anexados em atividades específicas no ambiente de aprendizagem, análogo ao ocorrido nos projetos em grupo.

### *4.2.2. Linguagens Portugol Studio e Java*

Até a primeira prova os alunos devem usar uma pseudo-linguagem de programação chamada Portugol Studio (lite.acad.univali.br/portugol), na qual os códigos são escritos em português estruturado, uma linguagem muito próxima à linguagem nativa do aluno. A partir da 4ª semana de aula é introduzida a linguagem de programação Java. A Figura 1 apresenta um código em Portugol Studio para ler dois números inteiros e imprimi-los em ordem crescente. A Figura 2 apresenta o código correspondente na linguagem Java. O código escrito em Java é mais complexo, como relatado em Kose & Deperlioglu (2012), comparado ao escrito em Portugol Studio.

```
programa {
     funcao inicio() {
          inteiro x, y
          escreva ("Digite um n´mero inteiro x:")
          leia (x)
          escreva ("Digite um n´mero inteiro y:")
          leia (y)
          se (x>y)
               escreva ("x > y:",x,y)
          senao se (x<y)
               escreva ("y > x:",y,x)
          senao
               escreva ("x = y:",x)
     }
}
```

*Figure 1. Código de um programa escrito na linguagem Portugol Studio.*

```java
package REC;
import java.util.Scanner;
public class ex01 {
    public static void main(String[] args) {
        int x, y;
        System.out.println("Digite um número inteiro x: ");
        x = (new Scanner(System.in)).nextInt();
        System.out.println("Digite um número inteiro y: ");
        y = (new Scanner(System.in)).nextInt();
        if (x > y) {
            System.out.print("x > y:" + x + ", " + y);
        } else if (y > x) {
            System.out.print("y > x:" + y + ", " + x);
        } else {
            System.out.print("x = y: " + x);
        }
    }
}
```

*Figure 2. Código de um programa escrito na linguagem Java.*

Posteriormente ao Exam1, o aluno que optar em entregar os códigos em Portugol Studio tem o seu conceito limitado a B. Foi percebido que este EP é útil para os alunos do BCT que desejam fazer cursos que não tem a computação como atividade principal, por exemplo, Biologia, Química ou Gestão. Estes alunos, geralmente, sentem dificuldades, ou falta de interesse, em utilizar uma linguagem mais complexa como Java, optando por Portugol Studio, mesmo tendo o conceito limitado a B. Como exemplo, na prova de recuperação de 2017.2 em IP-BL, dos 15 alunos que fizeram a primeira questão valendo 20% do conceito final, apenas 5 fizeram em Portugol Studio e tiveram conceitos B, B-, B-, C e D. Já na terceira questão, mais difícil e valendo 40%, somente 12 alunos entregaram e destes, 4 entregaram em Portugol Studio, com um conceito B e três conceitos F. Dos alunos que entregaram em Portugol Studio, apenas um ficou com conceito final F, os demais ficaram com conceito D. Os conceitos finais dos alunos que fizeram a prova de recuperação foram: um aluno com B, dois com C, nove com D e três com F.

### *4.2.3. Provas unificadas e individuais*

As avaliações Exam1, Exam2, SUB (prova substitutiva, para quem faltou Exam1, Exam2 ou Project) e REC (prova de recuperação) foram definidas sempre com 3 questões cada, de forma unificada, sendo cada questão de um nível de dificuldade: simples, média e complexa. Além disso, cada prova é única para cada aluno usando um banco de questões. Cada questão possui 4 ou 5 variações, conforme o número de alunos de uma turma, para garantir a individualidade de provas. Por exemplo, calcular o valor máximo/mínimo dos elementos pares/ímpares de um vetor de números inteiros, possibilita fazer 4 variações de uma questão para validar as competências de laços e condicionais. Assim, uma variação de cada questão é sorteada para cada aluno. Isso diminui possíveis plágios nas avaliações.

### *4.2.4. Correções unificadas*

No Exam1, cada questão é impressa em uma folha de papel frente e verso (geralmente o aluno consegue resolver a solução da questão na própria folha). Isso facilita a distribuição das provas para o professor que vai corrigir uma questão de todos os alunos de IP-BL.

Nas demais avaliações, o aluno deve submeter as soluções pelo Tidia-ae: códigos em Portugol Studio (arquivo .por) ou em Java (arquivo .java). Nestas avaliações, novamente um professor corrige uma única questão de todos os alunos de IP-BL.

### *4.2.5. Unificação para fechamento de conceitos*

Como o Tidia-ae não suporta o cálculo final de conceitos (letras), a partir dos critérios estabelecidos para IP-BL foi necessário criar uma planilha no Google Drive para gerenciar os vários conceitos atribuídos aos alunos, para contabilizar a sua participação no curso (>25% de não entregas significa conceito final F) e principalmente para o cálculo do conceito final. Para isso, foi feita uma conversão de conceito para nota (semelhante ao utilizado pela UFABC para o cálculo do Coeficiente de Rendimento (Grade Point Average) do aluno CR (GPA) (goo.gl/ZdpAhx, em português), como segue: A = 4; B = 3; C = 2; D = 1; F = 0; e conceito O = 0.

Em IP-BL, as variações +/- de um conceito significam decimais somando ou subtraindo o CR. Por exemplo, A-=3.8 e B+=3.2. Veja na Tabela 2 um exemplo de conversão de conceito para nota. Estes sinais +/- serão removidos no lançamento do conceito final do aluno no sistema acadêmico.

*Table 2. Tabela de conversão de conceito para nota.*

| A+ | A | A- | B+ | B | B- | C+ | C | C- | D+ | D | D- | E | F |
|---|---|---|---|---|---|---|---|---|---|---|---|---|---|
| 4 | 4 | 3.8 | 3.5 | 3 | 2.8 | 2.5 | 2 | 1.8 | 1.5 | 1 | 0.5 | 0 | 0 |

Agora, como atribuir o conceito final de uma prova, onde cada questão tem um peso diferente? Anteriormente, cada professor ficava responsável por atribuir esse conceito final, de forma subjetiva, após uma sugestão de cálculo usando os valores da Tabela 2 e os pesos por questão, veja Tabela 3. Nas duas últimas ofertas de IP-BL, esta tabela foi ajustada para cada avaliação, para todos os alunos e é utilizada de forma automática.

*Table 3. Tabela de conversão de nota para conceito.*

| ~CR | 0 | 0.8 | 1 | 1.5 | 1.8 | 2 | 2.5 | 2.8 | 3 | 3.4 | 3.75 | 3.9 | 4 |
|---|---|---|---|---|---|---|---|---|---|---|---|---|---|
| % | 0 | 0.4 | 0.42 | 0.45 | 0.48 | 0.5 | 0.6 | 0.65 | 0.7 | 0.75 | 0.8 | 0.85 | 1 |
| | F | D- | D | D+ | C- | C | C+ | B- | B | B+ | A- | A | A |

Por exemplo, considerando as três questões de uma prova com os respectivos pesos: Q1=25%, Q2=35% e Q3=40%. Se um aluno obteve A, B- e D+, respectivamente, nestas questões, teremos: Q1*25% + Q2*35% + Q3*40% = 4*25% + 3.2*35% + 1.2*40% = 2.6 = CR.

Este CR=2.6 está entre C+ e B-, considerando a Tabela 3. Cada professor pode interpretar de uma forma a atribuição do conceito final, deixando uma avaliação subjetiva e às vezes injusta. Como cada prova é única, com o mesmo grau de dificuldade, com critérios unificados nas correções de cada questão, seria justo também não deixar essa responsabilidade subjetiva de atribuição do conceito final para cada professor. Assim, a atribuição do conceito final é realizada ajustando os pesos da Tabela 3, para cada avaliação, e esse ajuste vale para todos os alunos. Na Seção 5 serão exemplificados os resultados desse EP.

### 4.3. Ferramentas de apoio

Para auxiliar nas avaliações, foram desenvolvidos scripts na linguagem de programação Python (python.org). O propósito dos scripts é a complementação da funcionalidade do VLE para gestão do processamento das avaliações e divulgação dos resultados. Tarefas repetitivas que são realizadas pelo professor durante a correção de uma avaliação, como a compilação de programas codificados pelo aluno ou a redação de feedback de erros e acertos, foram parcialmente automatizados. Todos os scripts a seguir estão disponíveis livremente para adapatações em vision.ufabc.edu.br/courses/ip-bl/scripts, com tutoriais de uso.

**script1copy:** compila o código submetido pelo aluno no VLE, agrupando em uma única pasta. Isso facilita a correção pelo professor e também a verificação de plágio, usando por exemplo JPlag (Prechelt & Malpohl, 2002) ou o trabalho de Liu et al. (2015). O professor deve atribuir o conceito e os erros no próprio nome do arquivo, por exemplo, "B-,2,4,5_RAaluno.pdf". Estes códigos de erros 2, 4 e 5 serão detalhados ao aluno no passo seguinte;

**script2email:** envia um email para cada aluno de forma automática, após a atribuição manual do professor/tutor do conceito e dos erros de cada questão/atividade. Este script também cria um arquivo CSV com todos os alunos e seus conceitos;

**script3MCTestIP:** Em toda avaliação presencial de IP-BL foi utilizado o software MCTest (Zampirolli et al., 2016) para a geração automática de provas individuais por alunos. Este software foi copiado, adaptado e renomeado para **script3MCTestIP**. Para cada avaliação é gerado de forma automática um arquivo PDF contendo todas as provas de uma turma, identificadas pelo nome do aluno e sua matrícula, como número e também como código de barras, veja Figura 3. As questões são sorteadas de um banco de questões, como já descrito;

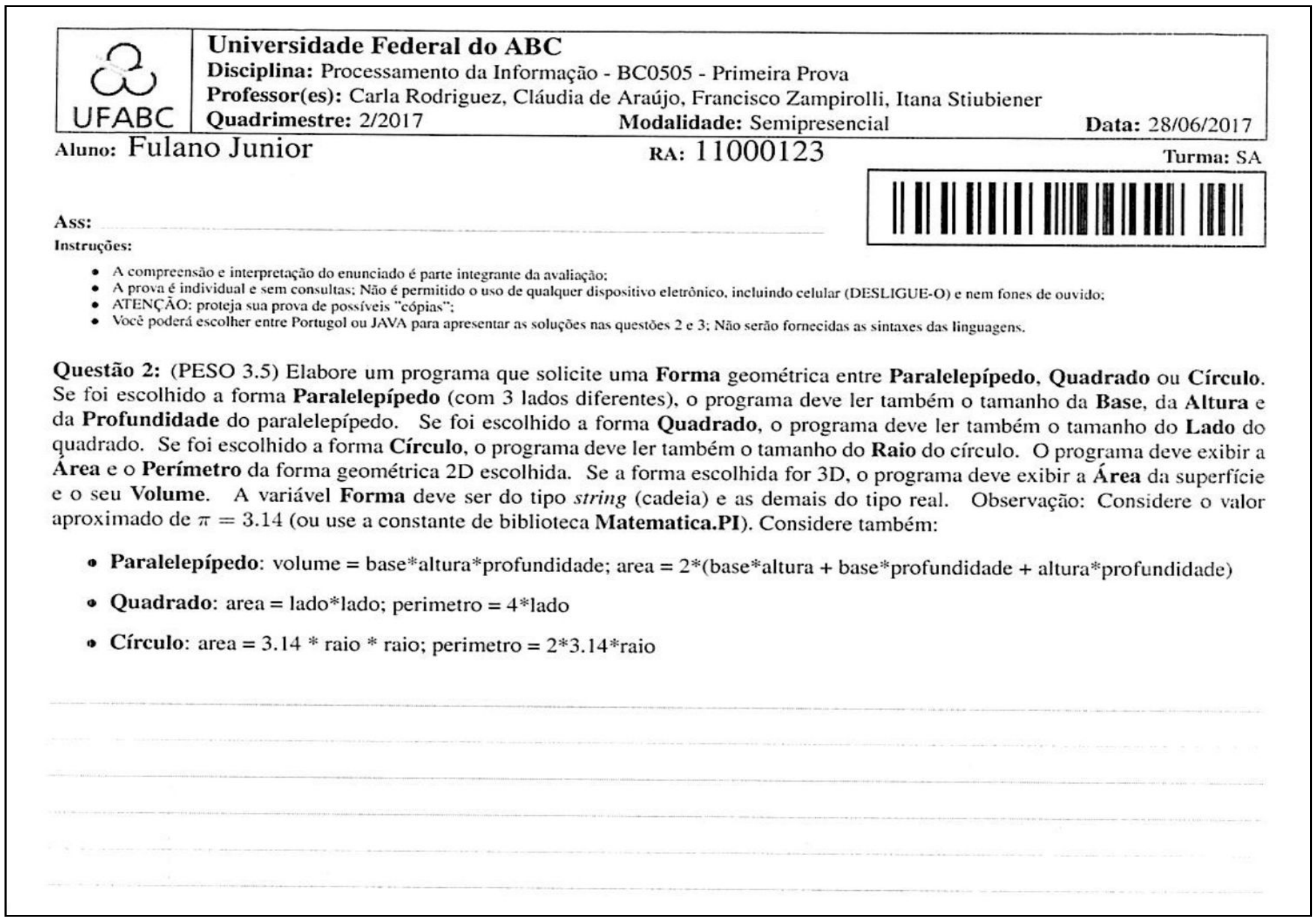

**Universidade Federal do ABC**
**Disciplina:** Processamento da Informação - BC0505 - Primeira Prova
**Professor(es):** Carla Rodriguez, Cláudia de Araújo, Francisco Zampirolli, Itana Stiubiener
UFABC **Quadrimestre:** 2/2017 **Modalidade:** Semipresencial **Data:** 28/06/2017

**Aluno:** Fulano Junior **RA:** 11000123 **Turma:** SA

**Ass:**

**Instruções:**

- A compreensão e interpretação do enunciado é parte integrante da avaliação;
- A prova é individual e sem consultas; Não é permitido o uso de qualquer dispositivo eletrônico, incluindo celular (DESLIGUE-O) e nem fones de ouvido;
- ATENÇÃO: proteja sua prova de possíveis "cópias";
- Você poderá escolher entre Portugol ou JAVA para apresentar as soluções nas questões 2 e 3; Não serão fornecidas as sintaxes das linguagens.

**Questão 2:** (PESO 3.5) Elabore um programa que solicite uma **Forma** geométrica entre **Paralelepípedo**, **Quadrado** ou **Círculo**. Se foi escolhido a forma **Paralelepípedo** (com 3 lados diferentes), o programa deve ler também o tamanho da **Base**, da **Altura** e da **Profundidade** do paralelepípedo. Se foi escolhido a forma **Quadrado**, o programa deve ler também o tamanho do **Lado** do quadrado. Se foi escolhido a forma **Círculo**, o programa deve ler também o tamanho do **Raio** do círculo. O programa deve exibir a **Área** e o **Perímetro** da forma geométrica 2D escolhida. Se a forma escolhida for 3D, o programa deve exibir a **Área** da superfície e o seu **Volume**. A variável **Forma** deve ser do tipo *string* (cadeia) e as demais do tipo real. Observação: Considere o valor aproximado de $\pi = 3.14$ (ou use a constante de biblioteca **Matematica.PI**). Considere também:

- **Paralelepípedo**: volume = base*altura*profundidade; area = 2*(base*altura + base*profundidade + altura*profundidade)
- **Quadrado**: area = lado*lado; perimetro = 4*lado
- **Círculo**: area = 3.14 * raio * raio; perimetro = 2*3.14*raio

*Figure 3. Exemplo de questão dissertativa do Exam1, após escanear.*

**script4corrections**: lê um arquivo PDF de provas escaneadas de uma turma; este script cria subpastas, por exemplo, P1scannerQ1, P1scannerQ2 e P1scannerQ3, conforme a questão identificada no código de barras; além disso, cria nestas subpastas arquivos PDF para cada aluno, um arquivo para cada questão, contendo uma questão escaneada frente e verso. Esses arquivos são ordenados pelo número de matrícula do aluno. Assim, o professor pode usar o mesmos critérios utilizados no **Exam1** para atribuir um conceito e apontar os erros na questão de cada aluno, por exemplo, "B-,2,4,5_RAaluno.pdf". Finalmente, é possível usar o **script2email** para enviar por email para o aluno a correção da questão, juntamente com o feedback dos erros.

## 5. RESULTS AND DISCUSSIONS

Esta seção destaca inicialmente alguns problemas enfrentados nas ofertas de IP-BL, que de certa forma, coincidem com os problemas de ausências de um processo unificado de avaliação em IP-Class. São descritas as soluções adotadas para lidar com cada problema e também novas sugestões para futuras ofertas. Além disso, é apresentada uma análise dos desempenhos dos alunos de IP-BL e IP-Class desde o início de cada oferta.

### 5.1. Problema na atribuição subjetiva do conceito final

A atribuição de conceitos pelo professor, de forma subjetiva, pode ocasionar injustiças, possibilitando, por exemplo, que um aluno com melhor desempenho obtenha conceito inferior ao de um aluno com desempenho pior. A Tabela 4 apresenta um cenário o qual todos os alunos deveriam fazer a REC por estarem reprovados por faltas, porém:

- student0, obteve conceito D no Exam1 e no Exam2 e F nos Projects e Activities, configurando um CR=0.7. Pelas regras, na ocasião, CR = 1.0 representava CbREC = F (conceito antes da avaliação de recuperação). Porém o professor deste aluno decidiu que ele deveria ser aprovado com conceito D;
- Nesta turma, de 39 alunos com CR entre 0.7 e 1.17, 7 deles obtiveram conceito D e optaram por não fazer a REC; dos 32 que obtiveram conceito F apenas 16 fizeram a prova REC e desses apenas 7 alunos foram aprovados;
- Os outros 9 alunos estão representados na Tabela 4 (student1 a student9); Note que student6, student7 e student8, mesmo obtendo conceito D na REC foram reprovados (CF = F); Ressalta-se que Exam2 e a REC são avaliações que abrangem todo o conteúdo do curso.

Assim, student0 mesmo tendo desempenho pior (CR = 0.7), foi aprovado e student6, student7 e student8 foram reprovados mesmo tendo obtido conceito D em REC. Este é apenas um exemplo de injustiças cometidas quando o cálculo do conceito final ocorre de forma subjetiva. Situações similares podem ocorrer nas mudanças subjetivas de conceitos entre D e C, C e B e de B e A.

*Table 4. Tabela com os conceitos de 9 alunos que fizeram uma REC.*

| weight | 30% | 15% | 15% | 40% | | | | |
|---|---|---|---|---|---|---|---|---|
| student | Exam1 | Activ. | Project | Exam2 | CR | CbREC | REC | CF |
| **student0** | **D** | **F** | **F** | **D** | **0.7** | **D** | - | D |
| student1 | D | B- | D- | - | 0.8 | F | F | F |
| student2 | D | B | D- | F | 0.83 | F | F | F |
| student3 | D | B- | D | - | 0.87 | F | F | F |
| student4 | D | B+ | D | F | 0.98 | F | F | F |
| student5 | D | B+ | D | F | 0.98 | F | F | F |
| **student6** | **C-** | **B+** | **F** | **F** | **1.07** | F | **D** | **F** |
| **student7** | **C-** | **B** | **D-** | **F** | **1.07** | F | **D** | **F** |
| **student8** | **D+** | **B+** | **D** | **F** | **1.13** | F | **D** | **F** |
| student9 | D | B- | B | F | 1.17 | F | F | F |

A seguir, são descritas as soluções adotadas para lidar com o problema da subjetividade na atribuição do conceito final (antes ou após a REC).

Foi observado também em uma outra turma de IP-BL, com professores muito generosos nos critérios de avaliação, onde o aluno que tirava CR >= 0.5 era suficiente para passar. Além disso, foi utilizado uma prova REC com critérios generosos de quem podia participar, onde todos os alunos podiam fazer a prova, foi também considerado a maior nota como a final. Claramente, o índice de aprovação nesta turma foi alto (81.45%)!

### *5.1.1. Mudanças implementadas nas turmas de IP-BL em 2017.1 e 2017.2*

Na planilha compartilhada entre os professores no Google Drive foi retirado a coluna de conceito atribuído pelo professor e foram ajustados os pesos nas fórmulas de cada uma das avaliações: Activities, Exam1, Project, Exam2, CbREC, e CF. Esses pesos são usados automaticamente para todos os alunos. Por exemplo, no Exam2 da turma de 2017.2 diminuímos o CR de A de 3.75 para 3.5 para todos os alunos, quando foi notado que isso era justo e adequado, analisando-se todos os casos com CR>=3.5. Para essa análise, foi necessário ordenar todos os alunos pelo CR para vislumbrar os casos similares e então fazer os ajustes necessários nos pesos da Tabela 3.

### *5.1.2. Novas sugestões de melhorias*

Tentando usar os critérios que já estão nas orientações iniciais, na primeira aula presencial, onde os alunos que estiverem melhorando nas avaliações presenciais serão beneficiados nos arredondamentos do conceito final da seguinte forma: incorporar esse cálculo nas avaliações antes da REC e na avaliação final. Por exemplo, somar no conceito final de cada aluno uma porcentagem de Prova2-Prova1>0. Por exemplo, se um aluno tirou Prova1=F=0 e Prova2=A=4, então Prova2-Prova1=4 e no cálculo somamos 0.4 na média final desse aluno. Outro exemplo de acréscimo ao conceito final do aluno é dar um bônus para quem fez muitas atividades com bons conceitos. Desta forma, somar por exemplo 20% da média dos conceitos das atividades realizadas (#ativ feitas / #ativ totais * 20%) no conceito final. Por exemplo, se um aluno fez todas as atividade e ficou com A de média, acrescenta 0.2 no conceito final (essas regras também devem estar na aula inaugural).

Permitir ao professor somente a escrita dos conceitos de cada questão de Exam/Project (de preferência usando o **script2email)** para minimizar erro de digitação. Como normalmente são centenas de alunos, redigitar o conceito e feedback seja na planilha, seja no VLE pode gerar erros e discrepâncias. Ajustar os pesos após ordenar os dados pelo CR e fazer os ajustes nos pesos conforme a Tabela 3. Dessa forma, não haverá mais ajuste manual, individual e subjetivo por cada professor, minimizando as injustiças.

Além disso, seria recomendável que todos os professores utilizassem o mesmo critério na avaliação do conceito final. Por exemplo, recentemente tivemos um questionamento legítimo de um aluno que relatou o cenário apresentado na Tabela X.

*Table X. Tabela com os conceitos de dois alunos para relatar uma reclamação de um aluno.*

| | Prova1 | Ativ. | Projeto | Prova2 | %faltas | REC | MF |
|---|---|---|---|---|---|---|---|
| aluno1 | B | A- | A | C | 10% | | B |
| aluno2 | D | B- | C | C | 40% | A | B |

Pelas regras da disciplina, este aluno2 teve que fazer a prova de recuperação, pois tinha muitas atividades não entregues. Nesta prova, o aluno teve o conceito A, ficando com média final B, igual ao aluno1. A Prova2 envolve todo o conteúdo da disciplina, assim como a REC. Observe que o aluno2 conseguiu recuperar o conceito ruim na Prova2 obtendo A na REC, demonstrando total conhecimento do conteúdo da matéria. O aluno1 não teve a mesma oportunidade de fazer a REC pelas regras da disciplina. Neste caso específico, qual seria a forma mais justa de avaliação?

Como o objetivo primordial de um professor é fornecer a maior quantidade possível de conhecimento para o aluno e avaliar esse conhecimento adquirido. Não seria melhor liberar a prova de recuperação para todos os alunos e considerar o maior conceito entre a média da MF antes da REC e o conceito da REC? Ou seja, liberar a REC para todos os alunos que quiserem e se conseguir recuperar alguma deficiência, considerar também esta melhora? Mas isso não pode ser opcional, onde cada professor decide o que fazer, pois vai acarretar injustiças, como as descritas anteriormente.

Considerando este cenário onde qualquer aluno pode fazer a prova de recuperação para melhorar o conceito final, não faz sentido ter uma prova substitutiva, para os alunos que faltaram Prova1, Projeto ou Prova2, por motivo de doença, por exemplo. Basta este aluno fazer a prova de recuperação e considerar o melhor caso, por exemplo, entre a média de Prova1, Projeto, REC, (se o aluno faltou a Prova2), e também da REC. Em média 5% dos alunos fazem a SUB (porque só 5% conseguem apresentar comprovante) e é necessário reservar uma semana das 12 do quadrimestre para a realização desta SUB. Eliminando esta prova, é possível aproveitar melhor esta semana, como oferecer uma aula presencial adicional.

### 5.2. Análise de desempenho

A Tabela 6 apresenta o desempenho dos alunos nas modalidades IP-BL (desde a primeira oferta em 2013) e IP-Class (desde 2009). A primeira coluna mostra o período letivo. A segunda coluna mostra a porcentagem de reprovação dos alunos de IP-BL, com o correspondente número total de alunos exibido na terceira coluna e o número de turmas na quarta coluna. A coluna 5 exibe a porcentagem de reprovação em IP-Class, com o correspondente número de alunos exibido na coluna 6 e o número de turmas de laboratório exibido na coluna 7 (lembrando que todas as turmas nesta modalidade são formadas com 30 alunos em média, porém, com o trancamento, o número médio de alunos diminui). Um resultado apresentado nesta tabela que chama a atenção é que a reprovação média de IP-BL (37.11%) é um pouco maior, mas próxima a de IP-Class (30.33%), corroborando com o trabalho de Castillo-Merino & Serradell-López (2014), apesar de ainda não ser possível garantir que todas as turmas usaram os mesmos critérios de ensino, aprendizagem e avaliação.

*Table 6. Desempenho dos alunos de IP-BL vs IP-Class desde 2009.*

| | **IP-BL** | | | **IP-Class** | | |
|---|---|---|---|---|---|---|
| **School term** | AVERAGE of %failure | SUM of total | COUNTUNIQUE of class | AVERAGE of %failure | SUM of total | COUNTUNIQUE of class |
| **2009.3** | | | | 25.25% | 856 | 34 |
| **2010.1** | | | | 30.82% | 107 | 4 |
| **2010.2** | | | | 37.08% | 217 | 9 |
| **2011.1** | | | | 29.14% | 1216 | 47 |
| **2011.3** | | | | 42.20% | 262 | 11 |
| **2012.1** | | | | 30.66% | 1207 | 45 |
| **2012.3** | | | | 59.95% | 220 | 11 |
| **2013.1** | 12.94% | 85 | 1 | 30.29% | 997 | 38 |
| **2013.2** | | | | 55.84% | 215 | 12 |
| **2014.2** | | | | 31.54% | 1005 | 37 |
| **2014.3** | 38.51% | 148 | 1 | 35.82% | 192 | 9 |
| **2015.1** | 29.95% | 217 | 1 | 24.11% | 869 | 32 |
| **2015.2** | 41.74% | 134 | 1 | 28.88% | 267 | 10 |
| **2015.3** | 44.94% | 145 | 1 | | | |
| **2016.1** | 39.02% | 147 | 1 | 20.62% | 958 | 36 |
| **2016.2** | 22.02% | 175 | 1 | 39.24% | 227 | 9 |
| **2016.3** | 43.47% | 243 | 1 | | | |
| **2017.1** | 29.75% | 185 | 1 | 24.00% | 1113 | 40 |
| **2017.2** | 28.88% | 162 | 1 | 34.85% | 220 | 7 |
| **Grand Total** | **37.11%** | **1641** | **10** | **30.33%** | **10148** | **391** |

Complementando essa Tabela 6, a Figura 4a mostra a média dos conceitos em porcentagem das 391 turmas de IP-Class, nos últimos 9 anos. Por exemplo, a média de conceitos A foi de 18.80% nas turmas de IP-Class, sem avaliações unificadas. As linhas em laranja mostram as variações dos conceitos em desvio padrão para mais e para menos sobre a média encontrada. É possível perceber uma grande variação dos valores, apesar de ser uma média sobre uma amostra de 391 turmas (um eventual valor distorcido, fora da curva, tende a perder a sua importância se a amostra for grande). Isto mostra que a dispersão é elevada e persistente.

A mesma análise foi realizada sobre os 7 turmas IP-BL de 2013.1 até 2016.2 conforme Figura 4b. Neste período a avaliação ainda não usava o formato unificado, e cada professor atribuia o conceito final utilizando seus próprios critérios. Percebe-se uma redução na dispersão das atribuições dos conceitos entre as turmas, provavelmente decorrente do uso do mesmo processo de ensino-aprendizagem e do mesmo material didático em todas as turmas de IP-BL, utilizando o VLE.

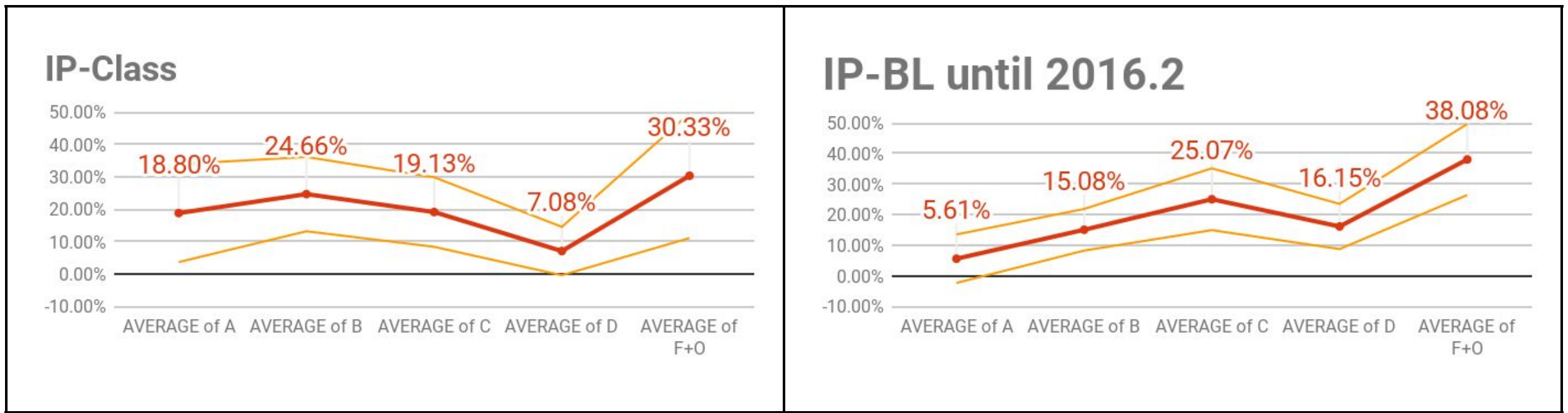

| (a) | (b) |
|---|---|

*Figure 4. (a) Desempenho dos alunos de IP-Class nos últimos 9 anos, com 391 turmas; (b) Desempenho dos alunos de IP-BL, de 2013.1 até 2016.2, com 7 turmas virtuais. Em vermelho apresenta a média desse desempenho; em amarelo representa o desvio padrão.*

Para comparações, a Figura 5a mostra a análise de IP-Class em 2017.1 e 2017.2, com 47 turmas, no mesmo período quando se adotou o formato de avaliação unificada descrito nesse trabalho aplicado em IP-BL. A Figura 5b mostra a análise sobre as turmas virtuais de IP-BL de 2016.3 até 2017.2. Apesar de ter-se muito menos dados, pois a análise é realizada sobre a amostra de apenas 3 turmas virtuais, é nítida a grande redução da dispersão da atribuição de conceitos entre as turmas. Nota-se também, comparando-se as figuras do IP-Class com os IP-BL, que a distribuição dos conceitos A, B, C, D e F+O (reprovação) estão mais coerentes nas turmas IP-BL do que nas turmas IP-Class, pois nestas últimas, prevalecem o Conceito "B" sobre os demais.

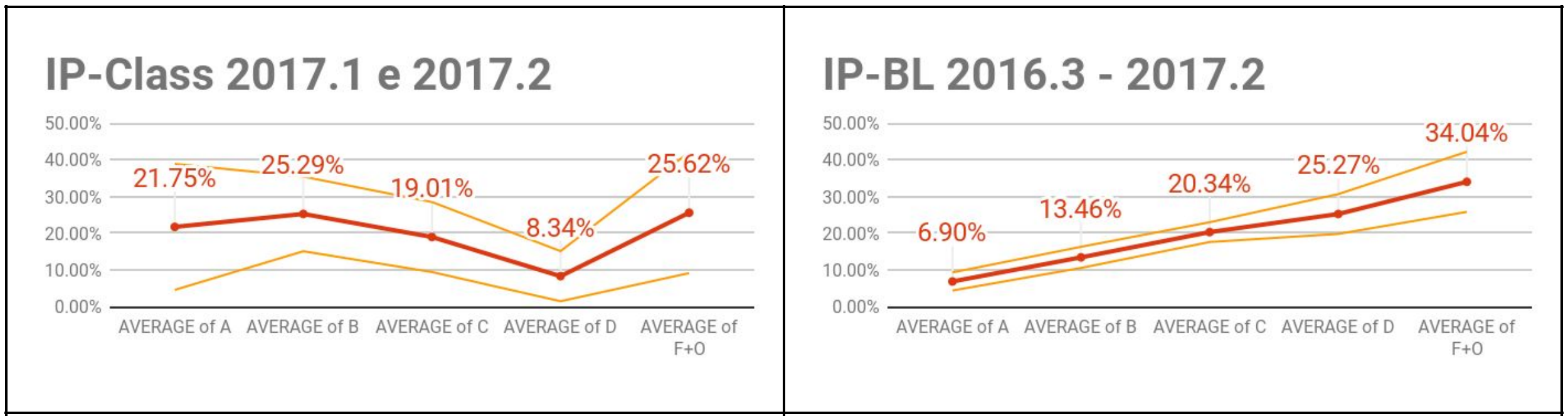

(a) (b)

*Figure 5. (a) Desempenho dos alunos de IP-Class em 2017.1 e 2017.2, com 47 turmas e 1333 alunos (não houveram turmas IP-Class em 2016.3); (b) Desempenho dos alunos de IP-BL, nas 3 turmas virtuais de 2016.3 até 2017.2, com 590 alunos, após as unificações das avaliações. Em vermelho apresenta a média desse desempenho; em amarelo representa o desvio padrão.*

A Tabela 7 apresenta o desempenho médio de todas as turmas de IP-BL e IP-Class, a partir de 2009, juntamente com o desvio padrão. Observamos que este desvio padrão é muito menor para o primeiro caso (veja última linha em negrito). Isso se justifica pois existe uma variedade grande nos critérios de avaliações adotados pelos professores na modalidade de IP-Class e também na modalidade IP-BL, antes do processo de unificação das avaliações/correções.

*Table 7. Desempenho médio e desvio padrão das modalidades de IP-BL e IP-Class.*

| | AVERAGE of A | AVERAGE of B | AVERAGE of C | AVERAGE of D | AVERAGE of F+O |
|---|---|---|---|---|---|
| **IP-Class** | 18.80% | 24.66% | 19.13% | 7.08% | 30.33% |
| ST | 15.12% | 11.46% | 10.73% | 7.47% | 19.23% |
| **IP-Class 2017.1 e 2017.2** | 21.75% | 25.29% | 19.01% | 8.34% | 25.62% |
| ST | 17.15% | 10.16% | 9.53% | 6.83% | 16.44% |
| **IP-BL until 2016.2** | 5.61% | 15.08% | 25.07% | 16.15% | 38.08% |
| ST | 7.94% | 6.81% | 10.11% | 7.39% | 11.63% |
| **IP-BL until 2016.3-2017.2** | 6.90% | 13.46% | 20.34% | 25.27% | 34.04% |
| ST | **2.47%** | **2.87%** | **2.69%** | **5.42%** | **8.18%** |

A Figura 6 mostra a variação do CR médio (average of GPA) calculado para cada turma de IP-Class. O CR médio é calculado considerando-se os valores da Tabela 2 para os conceitos A a F obtidos pelos alunos. O CR médio de uma turma (average of GPA for a class) é mais próximo de quatro quando há quantidade relativamente alta de alunos com conceitos A; quanto mais próximo de 0 indica que há considerável concentração de alunos reprovados. O gráfico da Figura 6 sinaliza, em cada quadrimestre, o menor e maior valor de CR de turma, além do intervalo de desvio padrão abaixo e acima do CR médio de turma. Em cada quadrimestre, os valores mínimo e máximo de CR médio destacam as diferenças avaliativas impostas por cada professor. Observe, a título de exemplo, os quadrimestres 2014.2 e 2016.1, quando a diferença entre o CR mínimo e máximo é maior que dois, que indica que a turma de pior desempenho obteve uma média de dois conceitos abaixo da turma de melhor desempenho. Também é possível observar que os quadrimestres em que houve maior quantidade de turmas IP-Class, e consequentemente maior quantidade de diferentes professores alocados, maior é o desvio padrão e a diferença entre mínimo e máximo.

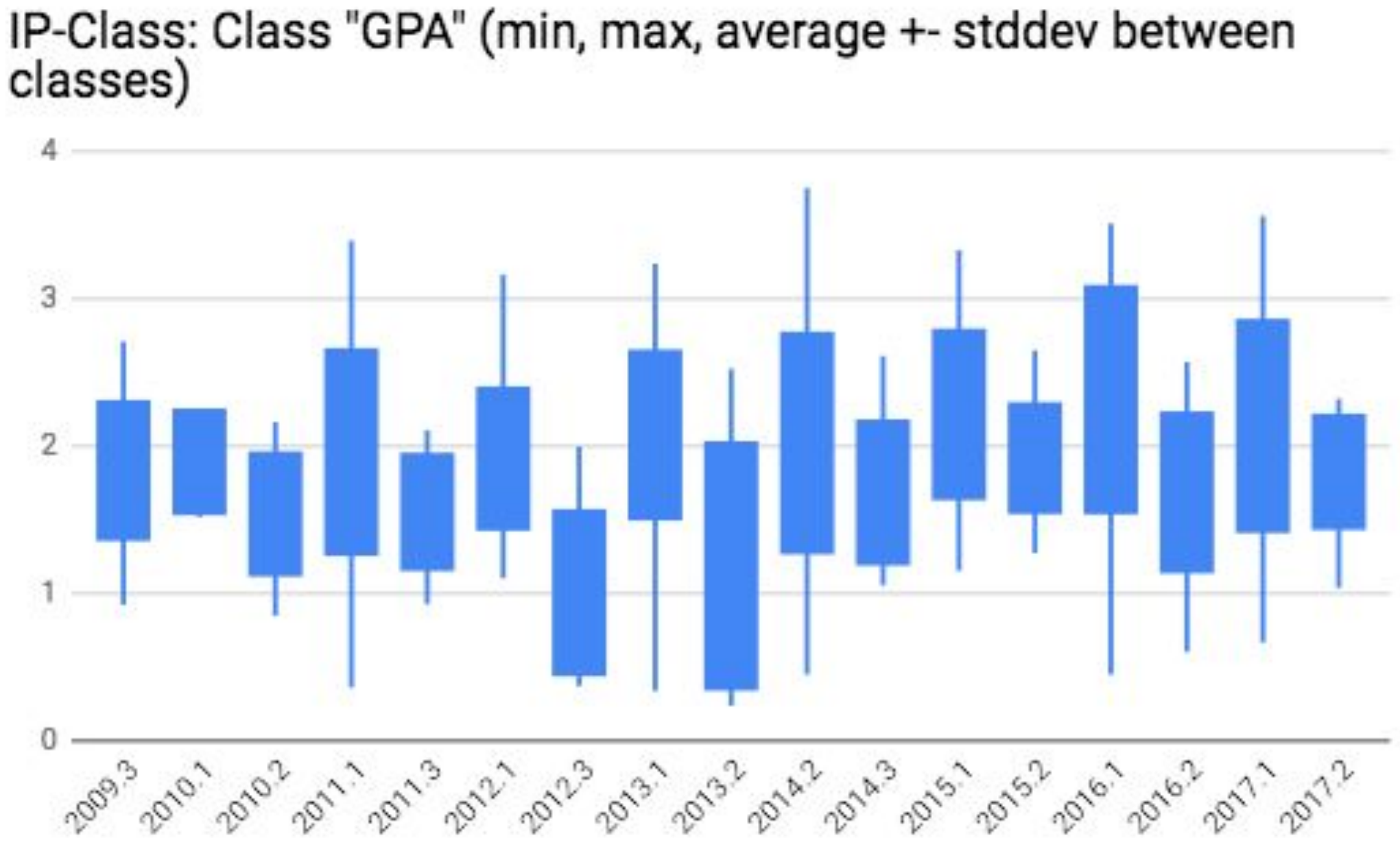

Figura 6: Dispersão das avaliações em IP-Class, com base no "GPA" de cada turma.

A Figura 7 aponta resultados de CR médio em cada turma de IP-BL. A quantidade de turmas de IP-BL em cada quadrimestre, em geral, é menor que a de IP-Class, porém cada turma de IP-BL costuma ter até duas ou três vezes mais alunos. Os quadrimestres 2013.1, 2014.3 e 2015.1 tiveram apenas uma turma virtual, assim, o valor pontual na Figura 7 nesses quadrimestres indica o CR médio dessas turmas únicas. Os quadrimestres 2016.1 em diante tiveram apenas duas turmas cada. É natural que a menor quantidade de turmas de IP-BL em cada quadrimestre produza desvios padrão relativamente menores aos observados em IP-Class na Figura 6. Entretanto, é notável que o CR médio das turmas IP-BL apresenta menor variação ao longo dos quadrimestres, comparativamente ao IP-Class (a média de IP-BL ficou entre 1 e 2, com exceção da primeira oferta em 2013.1). Essa menor variação, em parte, é devida à avaliação unificado aqui proposta e descrita.

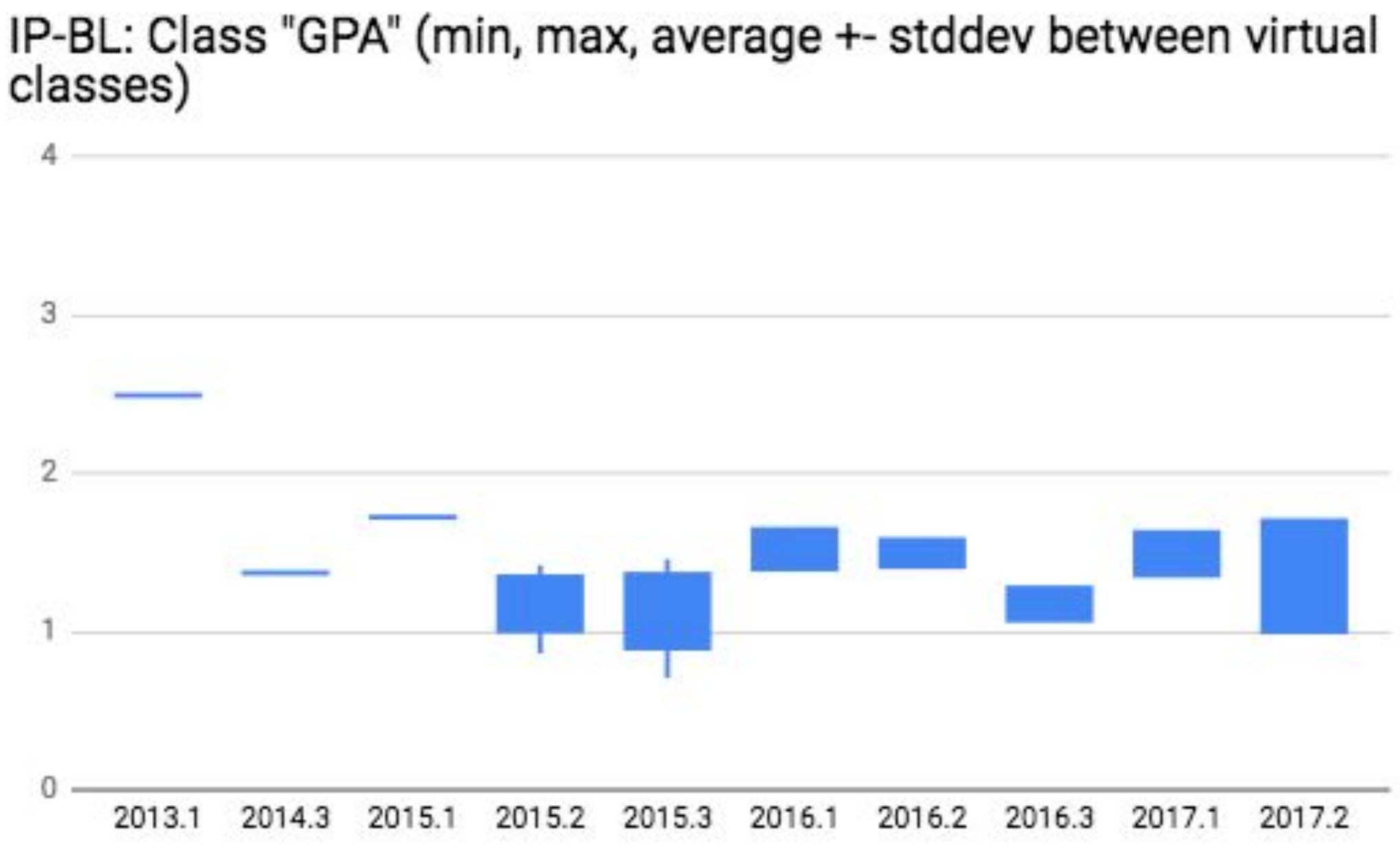

Figure 7: Dispersão das avaliações em IP-BL, com base no "GPA" de cada turma.

### 5.4. Número de reprovações anteriores

O perfil dos alunos de IP-BL apresenta em média 40% de alunos que tiveram pelo menos uma reprovação na mesma disciplina, como analisado a seguir (não foi possível obter esta medida para os alunos de IP-Class).

No início de cada quadrimestre, em IP-BL, é solicitado aos alunos o preenchimento de um formulário onde eles autodeclaram seu grau de conhecimento em programação e o número de reprovações anteriores em IP-Class e IP-BL. Na Tabela 8 são apresentados os números de alunos com 1, 2, 3 e 4 reprovações anteriores em IP-Class e/ou IP-BL, além do total de alunos que responderam o questionário. Esse número total de alunos é diferente do apresentado na Tabela 6, pois no início de cada quadrimestre ainda não ocorreu o ajuste de matrícula, no qual alguns alunos trancam e outros se matriculam após o início do quadrimestre.

Na Tabela 9 são apresentadas as porcentagens totais das reprovações em IP-Class e IP-BL. Ou seja, no último ano, 40.1% dos alunos matriculados em IP-BL tiveram alguma reprovação anterior. Além disso, apenas 16.4% desses alunos estão cursando IP-BL novamente, enquanto 32.7% tiveram reprovações em IP-Class. Com esses valores, podemos supor que IP-BL reprova menos que IP-Class. Por outro lado, pelas figuras da seção anterior, o número de alunos com conceitos A e B é menor em IP-BL, comparado com IP-Class, sugerindo alunos com mais dificuldades no aprendizado no primeiro caso. No entanto, sem avaliações padronizadas, em todas as turmas, essas inferências ficam comprometidas. Para se confirmar isso, seria necessário realizar mais experimentos nas duas modalidades usando o mesmos critérios de ensino, aprendizagem e avaliação.

Finalmente, a Tabela 9 apresenta 24.5% de alunos com reprovações em 2017.1, contra 43.3% (2016.3) e 52.6% (2017.2). Desta forma é possível observar que nos quadrimestres ideais para IP (quadrimestres com final .1), onde as turmas possuem alunos com menos reprovações anteriores, os alunos apresentam um desempenho melhor comparado com os outros dois quadrimestres (finais .2 e .3). Isso pode se confirmar também nos dados apresentados na Seção 5.2 nas duas modalidades (IP-Class e IP-BL).

*Table 8. Total de alunos com reprovações nos quadrimestre de 2016.3, 2017.1 e 2017.2.*

| | IP-Class | | | | IP-BL | | | | Total of Students |
|---|---|---|---|---|---|---|---|---|---|
| **Number of Failures** | **1** | **2** | **3** | **4** | **1** | **2** | **3** | **4** | |
| **2016.3** | 79 | 9 | 4 | | 22 | 6 | 3 | | **247** |
| **2017.1** | 25 | 5 | 1 | 1 | 25 | 3 | | | **184** |
| **2017.2** | 58 | 11 | 4 | | 32 | 7 | 1 | | **171** |
| **Grand Total** | **162** | **25** | **9** | **1** | **79** | **16** | **4** | | **602** |

*Table 9. Porcentagem de alunos com reprovações em IP-Class e IP-BL.*

| | IP-Class | IP-BL | Students with at least one failures |
|---|---|---|---|
| **2016.3** | 37.2% | 12.6% | 43.3% |
| **2017.1** | 17.4% | 15.2% | 24.5% |
| **2017.2** | 42.7% | 23.4% | 52.6% |
| **Grand Total** | 32.7% | 16.4% | **40.1%** |

## 5.5. Número de cancelamentos

Outro dado interessante é que o número de trancamentos. Nas 391 turmas de IP-Class tivemos uma média de cancelamentos de 13.58%. Já nas 10 turmas de IP-BL, tivemos uma média de 19.58% de cancelamentos. Porém, esta média diminuiu após uma resolução que limita para apenas um trancamento por quadrimestre, veja Figura 8.

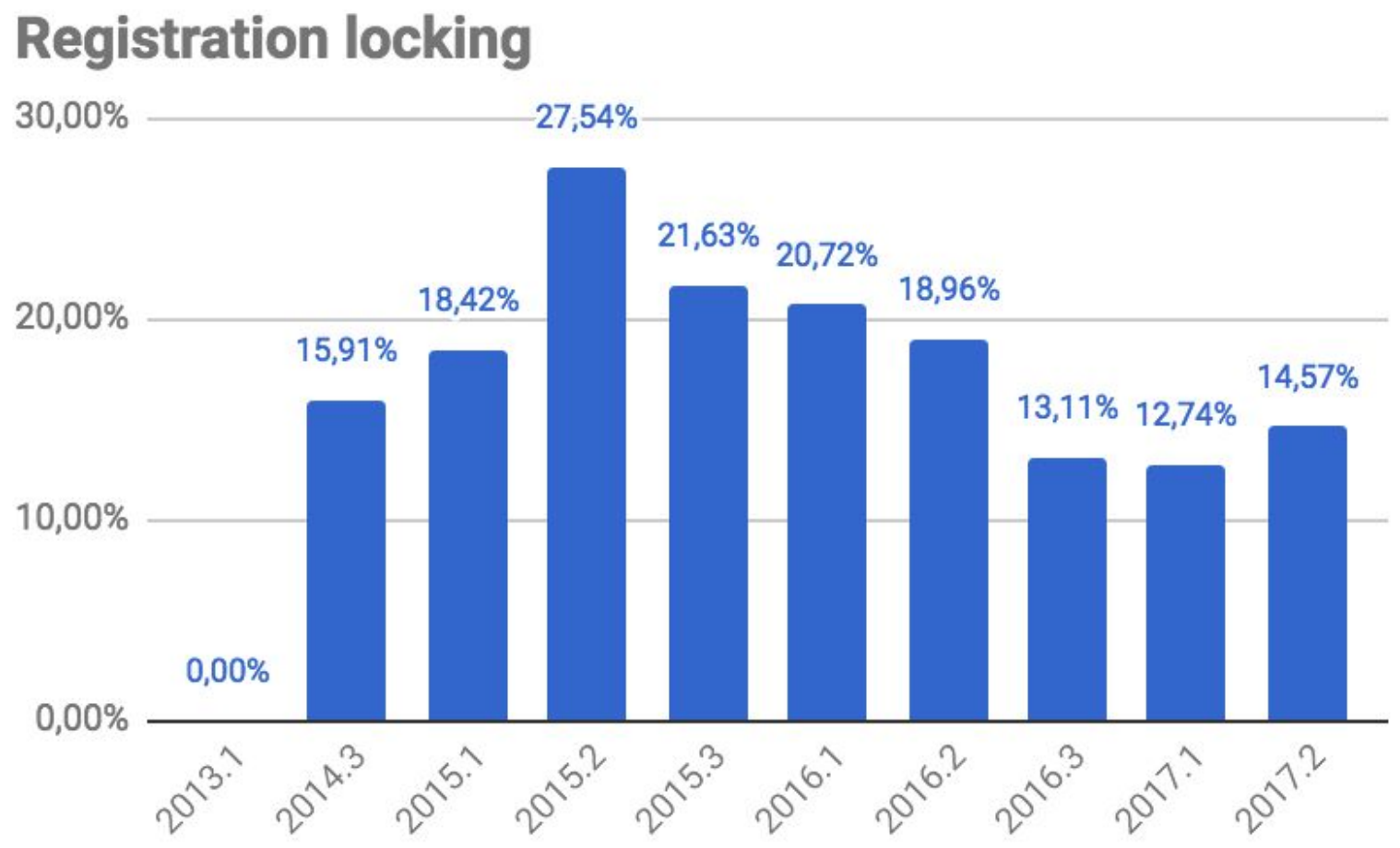

*Figure 8: Porcentagem de cancelamentos nas turmas de IP-BL.*

A Tabela Y apresenta uma alocação típica de um quadrimestre em IP-Class e IP-BL, antes do ajuste de matrícula (trancamento). A média de estudantes por laboratório em IP-Class neste quadrimestre de 2017.1 foi de 30.08. Após o ajuste de matrícula, a média de estudantes diminuiu para 27.83 por laboratório, uma redução de 7.48%. O aluno ingressa na UFABC no segundo quadrimestre do ano, assim IP tem o período ideal no primeiro quadrimestre no ano (final no primeiro ano de ingresso do aluno). No segundo e terceiro quadrimestres do ano temos uma demanda menor de alunos interessados em cursar esta disciplina, como constatado na Tabela 6.

*Table Y. Alocação típica de IP-Class e IP-BL no quadrimestre de 2017.1, antes do ajuste de matrícula (trancamento).*

| **Number of students by lab** | **Number of labs** | **Total of students** |
|---|---|---|
| 13 | 1 | 13 |
| 21 | 1 | 21 |
| 23 | 1 | 23 |
| 25 | 2 | 50 |
| 26 | 1 | 26 |
| 28 | 1 | 28 |
| 29 | 1 | 29 |
| 30 | 27 | 810 |
| 39 | 2 | 78 |
| 41 | 1 | 41 |
| 42 | 2 | 84 |
| **IP-Class-Total** | **40** | **1203** |
| **IP-BL** | | 201 |
| **Total** | | **1404** |

## 6. CONSIDERAÇÕES FINAIS E TRABALHOS FUTUROS

Este trabalho situa-se no contexto de cursos com múltiplas turmas e centenas ou milhares de alunos, que suscitam discrepâncias em avaliações, dificultando comparar desempenhos e a eficácia dos processos educativos. Este problema insere-se tanto na educação presencial como na educação a distância. De fato, a introdução de ICT na Educação tem propiciado construir instrumentos que podem melhorar a qualidade dos processos educativos.

Inicialmente, foram apontados problemas relacionados a falta de padrões em processos avaliativos a partir de turmas de Introdução a Programação na modalidade presencial (IP-Class), mas também constatados em turmas da mesma disciplina na modalidade blended learning (IP-BL). Propôs se então um processo para unificar as avaliações (aplicação e critérios) e relatou-se a aplicação do processo em 3 turmas de IP-BL.

Foram apresentados também 4 scripts para apoiar os professores na elaboração de provas únicas para cada aluno e também para ajudar no Evaluation Process (EP) de cada prova/atividade, com um professor corrigindo uma única questão de todos os alunos.

Finalmente, foi apresentado um comparativo de desempenho dos 1641 alunos das últimas 10 turmas de IP-BL ofertadas nos últimos 5 anos com os 10148 alunos na modalidade IP-Class (em 391 turmas nos últimos 9 anos). Foi constatado que o desempenho dos alunos nestas modalidades são equivalentes, com 37.11% de reprovações para IP-BL, um pouco pior que no modelo IP-Class, com 30.33%.

Os resultados mostram, no entanto, que após adoção do EP, nas turmas de 2016.3 a 2017.2, houve uma grande redução da dispersão da atribuição de conceitos entre as turmas, quando comparado a tanto a IP-Class quando a IP-BL antes da adoção de EP. Obviamente, os resultados não são conclusivos em virtude do número de turmas e alunos comparados.

A adoção desse EP como uma política da universidade depende, no entanto, da atualização das normativas da instituição e também da orientação e capacitação aos docentes. Entende-se que as propostas e soluções apresentadas neste trabalho podem ser adaptadas em outras instituições e também em outros cursos. Por exemplo, definir que as provas de recuperação devem ser em uma única semana, para todas as turmas de uma mesma disciplina. Assim, todos os alunos teriam o mesmo tempo para se preparar para a prova.

Apesar das ferramentas desenvolvidas facilitarem a automatização de tarefas manuais, entende-se que, como trabalhos futuros, seria recomendável incorporar todas as atividades e scripts utilizados no EP apresentados neste trabalho no VLE, eliminando a necessidade de uso de planilhas para fazer a gestão das avaliações e atribuição de conceitos finais, etc. A integração do VLE com ferramentas de correção automática de programas também é um trabalho futuro desejável.

Uma sugestão para acrescentar no Exam1 do curso, prova dissertativa em papel, seria incluir questões de múltipla escolha na prova. Nas atividades semanais, os alunos realizam testes com essas questões de múltipla escolha no próprio Tidia, que apresenta um feedback dos erros e a nota final, tudo automático. A proposta é escolher aleatoriamente 50% das questões que estão no banco de questões do Tidia para esses testes. Por exemplo, até a Prova 1 o aluno faz 5 testes de múltipla escolha, com 5 questões em cada teste e no banco de questões tem 50 questões classificadas com graus de dificuldade simples, média e complexa. A proposta é implementar um script que pega todas essas 50 questões do banco de questões do Tidia e converte para o formato utilizado pelo gerador de provas, apresentado neste trabalho como **script3MCTestIP**. Este gerador inclui em cada prova um teste de múltipla escolha, também com questões sorteadas aleatoriamente do banco. Por exemplo, cada prova pode ter 2 questões fáceis, 2 médias e 1 difícil. Assim, a Prova 1 teria estas questões de múltipla escolha mais 3 questões dissertativas. Desta forma, o aluno que fez as questões no Tidia, poderia ser beneficiado na prova (não sabemos quem realmente respondeu as atividades no Tidia). Essas questões de múltipla escolha da Prova 1 seriam corrigidas de forma automática, como descrito em Zampirolli et al. (2016).

Devido a alta taxa de de reprovações, que está em torno de 34% conforme resultados apresentados neste trabalho, entende-se que é necessário incorporar no processo de ensino-aprendizagem outras atividades de motivação e acompanhamento do aluno, como apresentado em Kuosa et al. (2016).